\documentclass[prl,twocolumn]{revtex4}

\newcommand{\calH}{{\cal H}}

\usepackage{bm}

\begin{document}

\title{Insulator, conductor and commensurability: a topological approach}

\author{Masaki Oshikawa}

\affiliation{ Department of Physics, Tokyo Institute of Technology \\
Oh-okayama, Meguro-ku, Tokyo 152-8551 JAPAN}

\date{December 15th, 2002}

\begin{abstract}
I discuss a topological relation of the conduction property of
a many-particle system on a periodic lattice at zero temperature
to the energy spectrum.
When the particle number per unit cell is
an irreducible fraction $p/q$, an insulator must
have $q$ low-lying states of energy $o(1/L)$ in one dimension
and of energy $o(1)$ in two dimensions, where $L$
is the linear system size.
\end{abstract}

\pacs{71.10.Fd 72.10.Bg 05.30.Jp}

\maketitle

An important question in the condensed matter physics is that
under what conditions the system becomes a conductor.
While we do not know, of course, the complete answer, there is a
characteristic of the system which may be naturally related to the conduction.
In a many-particle system on the periodic lattice, the known insulating
states such as band and Mott insulators can be understood
as a ``locking'' of the particles by the lattice.
It requires the particle density to be
commensurate with the lattice, namely
the filling factor (the number of particles per
unit cell of the lattice) to be an integer.
When the filling factor is not integral but is a (simple) rational
number, the particles may still be locked by spontaneously breaking
the translation symmetry and thus effectively enlarging the unit cell.
When a small number of particles are ``doped'' into the system,
they become mobile carriers, and we expect the system to be
gapless and conducting.
While this argument looks natural, it is far from trivial in the
strongly interacting systems. 
It is needed to clarify how far this physical expectation holds,
and what are the precise consequences.

In the recent several years, it has been pointed out that
the commensurability (in the sense of the particle number per unit cell)
is related to the topological nature of the system.
Extending the ideas by Lieb-Schultz-Mattis~\cite{LSM} (LSM)
and by Laughlin~\cite{Laughlin},
several statements on the energy spectrum
related to the commensurability are derived~\cite{OYA,comme}.
On the other hand, the naive expectation of the spontaneous breaking
of the translation symmetry apparently contradicts with
the ``Resonating-Valence-Bond'' (RVB) type
gapped quantum liquid states.
They have $1/2$ particle per unit cell, but there is no long-range order
which can be defined in terms of the local operator.
(For recent examples, see e.g. Refs.~\cite{MS-QDM,MSP-QDM,BFG}.)
However, this is still consistent~\cite{Bonesteel,Misg:degeneracy}
with the topological argument
as there is a ``topological degeneracy'' of the ground states
under the periodic boundary condition.
Moreover, the same underlying mechanism was
used in a topological proof~\cite{YOA,Luttad} of
the Luttinger's theorem~\cite{Luttinger}.
It was also applied
to the Fermi sea volume of a Kondo lattice,
which had been unamenable to rigorous analyses.

On the other hand, there still remain several fundamental questions,
including (but not limited to) the followings.
While the intuitive argument would relate the commensurability to the
conduction property, the relation has not
been discussed in the topological context.
Moreover, argument in Ref.~\cite{comme} for systems
in more than one dimension relies on a rather nontrivial
assumption that a finite gap does not collapse due
to an Aharonov-Bohm (AB) flux~\cite{AB}.
It is still necessary to clarify whether (or when) that assumption
can be justified.
In the present Letter, I shall address these questions by discussing
the Drude weight at zero temperature in view of the topological argument.

The classification of the insulator and conductor has been
discussed by several authors.
Drude weight represents the ``free acceleration'' of the current
by the electric field, and thus gives a criterion to define
the (perfect) conductor.
The Drude weight $D$ is defined by
\begin{equation}
 \sigma(\omega) = i \frac{D}{\pi}\frac{1}{\omega + i\delta}
+ \sigma_{reg}(\omega),
\label{eq:defD}
\end{equation}
where $\sigma(\omega)\equiv\sigma(\omega,\vec{q}=0)$ is the
zero momentum component of the complex AC conductivity,
$\sigma_{reg}(\omega)$ is assumed to be finite at $\omega \sim 0$,
and $\delta$ is a positive infinitesimal.
Kohn related the dependence of the ground state energy on the
AB flux to the Drude weight~\cite{Kohn}.
His argument has been further developed by several authors,
in particular by
Scalapino, White and Zhang~\cite{SWZ}.
In addition,
Resta and Sorella~\cite{RestaSorella} put forward
an alternative but deeply related criterion of an insulating state
which I will comment later.

Since the analyses on the Drude weight in Refs.~\cite{Kohn,SWZ}
and the topological argument on the commensurability are both
associated with the insertion of the AB flux, one may expect
they can be related to each other.
In order to establish a concrete connection,
let me first reformulate the Drude weight in the real time.
I consider a system defined on a finite $d$-dimensional periodic lattice 
of volume $V= L_x \times L_y \times \ldots \times L_d$ with periodic boundary
conditions, where $L_j$ denote the lengths in each directions.
The lengths and the volume are
measured in terms of the unit cell of the lattice.
Let the filling factor (particle number per unit cell of the lattice)
be $\nu = p/q$. I choose $L_x$ to be an integral multiple of $q$,
and all other $L_j$'s to be mutually prime with $q$.
In taking the thermodynamic limit $L_j \rightarrow \infty$ eventually,
$\nu$ is kept constant.

Suppose, for the time $t<0$, no electric field is applied and the system
was in the ground state. It can be chosen so that it is an eigenstate
of the $x$ component of the total momentum operator $P_x$.
For $0<t<T$ a small constant electric field $E$ is applied
uniformly throughout the system.
I then define the ``time-dependent Drude weight''
\begin{equation}
D(t) = \frac{\langle j(t) \rangle}{E t}
\label{eq:defDoft}
\end{equation}
where $j(t)$ represents the current in $x$-direction. 
Since the Drude weight represents the free acceleration
of the current by the electric field,
$D(t)$ converges to $D$ in the limit of $t\rightarrow \infty$.
Actually it can be proved,
provided that the linear response theory holds.
For a later convenience, I further define a weighted average $\bar{D}$
of $D(t)$ over the time domain $0<t<T$ by
\begin{equation}
\bar{D}  = \frac{2}{T^2} \int_0^T dt \; t D(t) .
\label{eq:defDbar}
\end{equation}
In the limit of large $T$, $\bar{D}$ is dominated by the long-time
asymptotic of $D(t)$, and thus also should converge to $D$.

One still has to be careful in using the present formulation, however.
For a given lattice model and a given (finite) electric field $E$,
$\langle j(t) \rangle$ cannot keep the $t$-linear behavior indefinitely.
In particular, in a lattice model,
$\langle j(t) \rangle$ would not exceed some finite upper limit.
This is a manifestation of the non-linear effects.
Generally, when $\langle j(t) \rangle$ becomes large,
non-linear effects are important and the linear response theory
breaks down.
In order to apply the linear response theory (so that
eqs.~(\ref{eq:defDoft}) and (\ref{eq:defDbar}) gives the
standard Drude weight),
the current $\langle j(t) \rangle$ should be sufficiently small.

Let us represent the electric field by the time-dependent vector
potential $A(t)$, which is uniform throughout the system,
in the $x$ direction. It is given by
$ A(t) = Et$.
In fact, this $A(t)$ is related to the AB flux $\Phi(t)$
which pierces through the ``hole'' of the torus.
$ \Phi(t) = L_x A(t)$.
The current $j(t)$ is given by
\begin{equation}
 j(t) = \frac{1}{V} \frac{\partial \calH}{\partial A} = \frac{L_x}{V}
      \frac{\partial \calH(\Phi)}{\partial \Phi},
\label{eq:joft}
\end{equation}
where $\calH(\Phi)$ is the Hamiltonian with the AB flux $\Phi$.
Since $\Phi$ is the function of time, Hamiltonian is also a
function of time.
\begin{equation}
\frac{d \calH}{d t} =  \frac{d\Phi}{dt}
\frac{\partial \calH(\Phi)}{\partial \Phi}
= E L_x \frac{\partial \calH(\Phi)}{\partial \Phi}.
\label{eq:dHdt}
\end{equation}
Combining eqs.~(\ref{eq:joft}) and~(\ref{eq:dHdt}), we obtain the
interesting relation
$j(t) = (1/EV) {d\calH}/{dt}$.
Now the averaged Drude weight $\bar{D}$ is given by
\begin{eqnarray}
 \bar{D} &=& \frac{2}{T^2 E^2 V}
  \int_0^T \langle \frac{d\calH}{dt} \rangle dt \nonumber \\
&=&
\frac{2 L_x^2}{\Phi(T)^2 V}
\big( \langle \calH \rangle_{t=T} - \langle \calH \rangle_{t=0} \big) .
\label{eq:DdH}
\end{eqnarray}
In the Schr\"{o}dinger picture, $\langle \calH \rangle_{t=T}$
should be understood as
$\langle \Psi(T) | \calH(\Phi(T)) |\Psi(T)\rangle$ where
$|\Psi(t)\rangle$ is the state vector at time $t$.
This formula is {\em exact} and does not rely on the adiabatic theorem.

If one takes $T$ to infinity and $\Phi(T)$ to infinitesimal,
the time evolution may become adiabatic
(however, see discussions below).
In this case,
the above formula reduces to the well known Kohn formula~\cite{Kohn}
\begin{equation}
D = \bar{D} = \frac{L_x^2}{V}
\frac{\partial^2 E_0(\Phi)}{\partial \Phi^2},
\label{eq:Kohn}
\end{equation}
where $E_0(\Phi)$ is the ground state energy as a function of the
AB flux ($\Phi$).

Scalapino, White and Zhang~\cite{SWZ} pointed out a subtlety in the
Kohn formula in particular for a gapless system in $d>1$.
Namely, many level crossing occur even for a very small AB flux
$\Phi$ in a large system, and the definition of
$\partial^2 E(\Phi)/\partial \Phi^2$ becomes actually rather subtle.
They offer some definitions which would give the correct Drude weight.
An alternative solution proposed here is to use
the real-time formulation~(\ref{eq:DdH}) instead of the Kohn
formula~(\ref{eq:Kohn}).
I fix the time interval $T$ (and thus the ``sweeping rate'' of $\Phi$)
and take the thermodynamic limit $L_j \rightarrow \infty $ first.
After taking the thermodynamic limit, I let $T \rightarrow \infty$.
There could be more and more level
crossings and the time evolution is not
necessarily adiabatic, as the system size is increased
with a fixed $T$.
With huge numbers of crossings in a large system, the final
state $|\Psi(T)\rangle$ may actually be very complicated~\cite{LZS}
but $\bar{D}$ is still well-defined.
By taking $T\rightarrow \infty$ limit after taking the thermodynamic limit,
$\bar{D}$ should converge to the standard Drude weight $D$.
This procedure in the real time corresponds to,
in the frequency space,
taking $\omega \rightarrow 0$ limit after taking the thermodynamic limit
with $\vec{q}=0$.

Now let me choose $\Phi(T)$ to be the unit flux quantum
$\Phi_0 = 2\pi/e^2$ where $e$ is the charge of the particle and
I set $\hbar=c=1$.
I note that, this choice corresponds to applying a small electric
field $E = O(1/(TL_x))$.
Even at $t=T$, the current is $j(T) \sim DET = O(1/L_x)$.
Thus, it can be arbitrarily small as one takes the limit of the
large system size $L_x$.
Therefore the linear response theory should be valid in a sufficiently
large system.
I now invoke the argument used in Refs.~\cite{AligiaOrtiz,comme,Luttad}.
Namely, as the Hamiltonian is translationally invariant for any
value of $\Phi$, the total momentum $P_x$ of the system does not change
during the process. It can be most simply seen in the Heisenberg
equation of motion $ dP_x / dt = i [ \calH, P_x ] = 0$.
However, at $t=T$, the Hamiltonian $\calH(\Phi_0)$ with a
unit flux quantum inserted is not identical to the original one $\calH(0)$,
although they must have the same spectrum.
The large gauge transformation
\begin{equation}
U \equiv e^{2\pi i \sum_{\vec{r}} x n_{\vec{r}}/L_x},
\label{eq:defU}
\end{equation}
where $n_{\vec{r}}$ is the particle density operator at site $\vec{r}$,
brings the Hamiltonian
$\calH(\Phi_0)$
back to the original form $\calH(0)$.
In the course, the total momentum $P_x$ of the system
is also changed by $ \Delta P_x = 2 \pi \nu L_y \times \ldots L_d$ because of
\begin{equation}
 U^{-1} T_x U = T_x e^{i \Delta P_x },
\end{equation}
where $T_x=e^{iP_x}$ is the lattice translation operator to $x$ direction.
The momentum change $\Delta P_x$ is different from $0$ for
an incommensurate filling ($\nu$ not an integer).
I note that this argument does not require the process
to be adiabatic. Even for a non-adiabatic time evolution and a
complicated final state $|\Psi(T)\rangle$, the
momentum is still exactly conserved before making the large gauge
transformation,
thanks to the translation invariance of the Hamiltonian.

Let me denote $E_{\Delta P_x}$
be the energy of the lowest excited state with total
momentum $\Delta P_x$.
Then $\bar{D}$ is bounded from below as
$\bar{D} \geq e^2 L_x^2 E_{\Delta P_x} / (4 \pi V)$.
Considering the case $\Phi(T) = 2\pi n /e^2$, we can obtain
similar bounds on $E_{n \Delta P_x}$ for $n =1,2,\ldots q-1$.

This expression suggests the nontrivial dependence of the transport
property and/or the energy spectrum on the aspect ratio~\cite{SWZ}.
In taking the thermodynamic limit, here I want to keep an isotropic
limit. Namely all the length $L_j$ are taken to be
of same order of magnitude $L$ and then $L$ is sent to the infinity.
In that limit, we obtain
\begin{equation}
 \bar{D} = D \geq \mbox{const.} \; L^{2-d} E_{n \Delta P_x},
\end{equation}
where $d$ is the dimensionality of the system.
This is the main result of the present letter.
Let me discuss its consequences for the systems in different
dimensions $d$ separately.

{\noindent $\mathbf d=1$}

If the system has an incommensurate filling and still an insulator ($D = 0$),
\begin{equation}
 E_{n \Delta P_x} = o(\frac{1}{L}),
\label{eq:d1bound}
\end{equation}
where $n=1,2,\ldots q-1$ and $o(1/L)$ means vanishing {\em faster} than $1/L$
as $L\rightarrow \infty$.
It means that, if there is a gap between the ground state(s) and
the excited states, there are $q$-fold (nearly) degenerate ground states
$|\Psi_0^{(k)}\rangle$ with different momenta $P_0+2 \pi k \nu $.
($k=0,2,\ldots,q-1$).
Intuitively, the system could acquire a gap by spontaneously breaking
the translation symmetry to become an insulator.
Then generally we expect $q$-fold degenerate ground states below the gap.
This is consistent with the above bound.
On the other hand, the result~(\ref{eq:d1bound}) does not rule out the
possibility of a gapless incommensurate insulator.
Still, the bound~(\ref{eq:d1bound}) requires that
its finite-size spectrum of such a system should not scale (entirely)
as $1/L$, although it is the case when
the system is described by a conformal field theory.

On the other hand,
Resta initiated the argument~\cite{RestaSorella,AligiaOrtiz} that
the phase of the ground-state expectation value of
eq.~(\ref{eq:defU})
\begin{equation}
z = \langle \Psi_0 | U | \Psi_0 \rangle,
\end{equation}
can be interpreted as the polarization which is well defined
only in an insulator.
Namely, the system is a conductor if $z=0$ and an insulator if $z\neq 0$
in the thermodynamic limit.
That this approach is rather closely related to the LSM argument
is pointed
out only recently by Nakamura and Voit~\cite{NakamuraVoit}.
Because of the LSM argument, $U |\Psi_0\rangle$ belongs to a
different momentum eigenvalue than
the ground state and thus $z=0$.
This naively implies that the system is always a conductor, which
is of course not true.
Observing this, Aligia and Ortiz~\cite{AligiaOrtiz} proposed to use
$z^{(q)} = \langle \Psi_0 | U^q | \Psi_0 \rangle$ instead of $z$,
for a fractional filling $\nu=p/q$.
This indeed fits with the picture of
spontaneously broken translation symmetry, which implies
the expectation value of the position operator to be multi-valued.
In such a case,
the physical (pure) ground state in the thermodynamic limit
should be taken as
$|\Phi^{(j)}\rangle \sim \sum_{k} e^{2\pi j k\nu}|\Psi_0^{(k)}\rangle$.
The original order parameter $z$ can be non-vanishing with respect
to $|\Phi^{(j)}\rangle$.
More analyses would be needed in this direction to examine
the implications.

In any case, the present result implies that it would be
more 'difficult' to make
an insulator out of a system with highly incommensurate filling
({\it i.e.} large $q$) as there are large number of restrictions.
Even when the system is conducting, requiring the Drude weight to be finite 
we obtain $E_{n\Delta P_x} = O(1/L)$ if the filling is incommensurate.
While this is nontrivial, the standard
LSM argument~\cite{LSM,OYA,YOA}
already gave this bound irrespective of the conduction property.
On the other hand, eq.~(\ref{eq:d1bound}) is a stronger result
which applies only to an insulator. 
The bound~(\ref{eq:d1bound}) certainly should not apply to
a conductor, as there are many counterexamples.

{\noindent $\mathbf d=2$}

If the system has an incommensurate filling and still an insulator
($D=0$),
\begin{equation}
 E_{n \Delta P_x} = o(1).
\label{eq:d2bound}
\end{equation}
Namely, they should vanish (at some rate, possibly very slowly)
in the thermodynamic limit.
In particular, when there is a finite gap between the ground states and
the excited states, the ground states must be $q$-fold degenerate.
On the other hand, a gapless insulator phase is also consistent with
the above bound.
This conclusion is essentially the same as obtained in
Ref.~\cite{comme} assuming that the gap does not close due to the
AB flux.
Here, that (questionable) hypothesis is replaced by the assumption
that the system is an insulator.
For a conductor, requiring the Drude weight to be finite
only yields a trivial bound $E_{n \Delta P_x} = O(1)$,
Unfortunately, no useful result can be obtained
for $d \geq 3$ with the present argument.

Straightforward generalization to the multi-species cases
is possible in parallel to Ref.~\cite{YOA}.
For example, let me define $\nu_{\uparrow}$ and $\nu_{\downarrow}$
respectively as up-spin and down-spin electron density,
and $\nu=\nu_{\uparrow} +\nu_{\downarrow}$.
One can consider the ``real'' electric field which couples equally
to electrons with either spin,
to derive similar constraints if $\nu$ is fractional.
For the half-filled Hubbard model, this does not give any constraint
as $\nu=1$ is an integer.
However, in this case one can still consider a fictitious electric
field which couples to up-spin electrons only.
Then applying the same argument as before, one can deduce the
bounds~(\ref{eq:d1bound}) and~(\ref{eq:d2bound}) respectively
for one and two dimensions, if the system is ``insulating''
{\em with respect to the up-spin electrons}.
For the standard Hubbard model in one dimension at half-filling,
the exact Bethe Ansatz solution shows that all the low-lying states
scale as $1/L$ for any value of the Hubbard interaction $U$.
This is actually consistent with the above result, because
for $U>0$ the system is insulating with respect to the (real) charge
but is a ``conductor'' in terms of spin.
For $U<0$, it is a ``spin insulator'' but is a charge conductor.
Therefore, the system does conduct ``number of up spin electrons''
for either signs of $U$.

To summarize, in this Letter I have discussed a relation among
the commensurability, the conduction property of the system,
and the finite-size energy spectrum, combining the ideas
from Refs.~\cite{Kohn,SWZ} and those from Refs.~\cite{LSM,Laughlin,OYA,comme}.
As a consequence, I have derived an upper bound for energies of some
low-lying states for an insulator with an incommensurate filling
in one and two dimensions. 
It is amusing to observe that several of very
fundamental concepts~\cite{LSM,Laughlin,Luttinger,AB,Kohn,RestaSorella}
in condensed matter physics, some of them
put forward about 40 years ago, are rather deeply connected with
each other, and to the simple notion of the commensurability.

As a final remark, I re-examine
the validity of the ``gap protection'' hypothesis made in Ref.~\cite{comme}.
Suppose the system has a finite gap (at zero AB flux) above the ground
state(s).
If the hypothesis is valid ({\it i.e.} the gap remains finite for any value
of the AB flux), the present argument concludes that the system is
an insulator ($D=0$).
Therefore, if there is a conductor ($D>0$) with a finite gap, 
it contradicts with the hypothesis.
According to Ref.~\cite{SWZ}, in the presence of the gap,
the superfluid weight $D_s$ is identical to the Drude weight $D$,
thus the conductor with a finite gap automatically means a superconductor.

In fact, in the presence of the gauge (Coulomb) interaction,
a superconductor generally has a finite mass gap, due to the 
Anderson-Higgs mechanism~\cite{AH}.
Thus, the gauge (Coulomb) interaction, being long-ranged, is enough
to invalidate the hypothesis.
If there is no long-range Coulomb interaction, a ``superconductor''
(or rather a superfluid) is generally associated to
off-diagonal long-range or quasi-long-range order, and thus
to gapless excitations.
Hence the hypothesis may still be valid for a systems with
short-range interactions (and hoppings).

Recently, Misguich {\it et al.}~\cite{Misg:degeneracy}
verified the ``gap protection'' hypothesis numerically
for the RVB spin liquid state in $d=2$.
They actually proposed an even stronger hypothesis (at least for the
spin liquid state in $d=2$) that the ground-state energy $E(\Phi)$
is independent of the AB flux $\Phi$ in the thermodynamic limit.
This ``flatness'' hypothesis in Ref.~\cite{Misg:degeneracy} may
be more naturally understood with respect to the conduction property.
Namely, if the groundstate is always below the gap, its energy
$E(\Phi)$ must satisfy $E(\Phi_0)=E(0)$.
Thus, if $E(\Phi)$ varies with $\Phi$, it must be convex at one
point while it is concave at another. 
According to eq.~(\ref{eq:Kohn}) the system must then change
its character from a superconductor to an insulator due
to the insertion of the AB flux.
Excluding such an unnatural behavior leads
to the ``flatness'' hypothesis in $d\leq 2$.
On the other hand, this argument does not support
the ``flatness'' hypothesis for $d \geq 3$.

It is a pleasure to thank Leon Balents, Claudio Chamon, Gregoire
Misguich and Masaaki Nakamura for very useful comments.
The present work is supported in part by Grant-in-Aid for Scientific
Research from MEXT of Japan.

\end{document}